\documentclass{css2016}

\usepackage{epsfig}

\newcommand{\PACS}{\MSC}

\title{Two fundamental cosmological laws \\
      of the Local Universe}

\author{Yurij V. Baryshev$^1$\\
\vskip 2mm {\small
$^1$
Astronomical Department
of the Saint Petersburg State University\\
Universitetskij pr.28,
Stary Peterhoff,
St. Petersburg, 198504\\
yubaryshev@mail.ru \\
}}

\abstract{
The Local Universe is the most detail studied part of the observable region of space 
with the radius R about 100  Mpc. 
There are two empirical fundamental cosmological laws
directly established from observations in the Local Universe
independently from cosmological theory:
first, the Hubble-Humason-Sandage linear redshift-distance law
and second, 
Carpenter-Karachentsev-deVaucouleurs density-radius power-law.
Review of modern state of these empirical laws and their cosmological significance 
is given. Possible theoretical interpretations of the surprising coexistence of both laws at the spatial scales from 1 Mpc to 100 Mpc are discussed.
Comparison of the standard space-expansion explanation of the cosmological redshift with possible global gravitational redshift model is given.
}



\begin{document}

\maketitle

\section{Introduction}
Cosmology as a physical science is based on observations, experiments
and theoretical interpretations.
Hubble 1937 \cite{hubble37} put forwarded
"The Observational Approach to Cosmology". It was developed later by
Sandage 1995a \cite{sandage95a} who  used the term "Practical Cosmology" to denote the observational study of "our sample of the Universe", which delivers possibilities for testing alternative initial hypotheses and main predictions of cosmological models.

Cosmology deals with a number of empirical
facts among which one hopes to find fundamental laws.
This process is complicated
by great limitations and even under the paradigmatic grip of any current standard cosmology.
One should distinguish between two kinds of cosmological
laws:
\begin{itemize}
  \item directly  measured empirical laws,
  \item logically inferred theoretical laws.
\end{itemize}

The empirical laws are directly measured relations between observable
quantities, which should be corrected for known selection and 
distortion effects.
The logically inferred theoretical laws
(theoretical interpretations) are made on the basis of an accepted
cosmological model, e.g. the standard or an alternative cosmological model.
Theoretical derivations utilize modern theoretical physics and even its
possible extensions, which can be tested by observations.

During one hundred years of intensive investigations
of the Local Universe (which can be defined as region
of space with radius $R$ about 100 Mpc )
two especially important cosmological
empirical laws were unveiled (see review in
\cite{baryshev-t12}, \cite{baryshev-t02}, \cite{baryshev94}):
\begin{itemize}
  \item the cosmological linear redshift-distance law $cz=HR$,
  \item the power-law correlation of galaxy clustering
  $\Gamma(r)\propto r^{-\gamma}$.
\end{itemize}
Here $R$ is the distance to a galaxy, $H$ is the Hubble constant, $r$ is the radius of test spheres around each galaxy,  $\Gamma(r)$ is the complete correlation
function (the conditional density)  and $\gamma$ is 
the power-law exponent.

The empirical laws, being based on repeatable observations, are independent of existing or future cosmological models.
However, the derived theoretical laws are valid only in the frame
of a specific cosmological model. Good examples are the empirical Hubble
linear redshift-distance ($z \propto R$) law and the derived theoretical 
space-expansion
velocity-distance ($V_{sp-exp} \propto R$) law within the Friedmann model.

An analysis of both empirical cosmological facts and theoretical initial assumptions
together with main logical inference in the frame of the standard and several alternative cosmological models is presented in our book
Baryshev \& Teerikorpi 2012 \cite{baryshev-t12}.
Below I concentrate on the significance for cosmology the redshift-radius and
density-radius empirical cosmological laws.

\section{Hubble-Humason-Sandage linear redshift-distance law}
The linear relation between cosmological redshift and distance to galaxies
was first established by Hubble 1929 \cite{hubble29} using distance estimations
for 30 galaxies at very small scales
$1 \div 10 $ Mpc, corresponding to redshifts $z < 0.003 $ or
spectroscopic radial velocities $v_{rad} < 1000$ km/s.

The extension of the linearity of the redshift-distance relation up to redshifts
about $z < 0.05$ or scales about 150 Mpc was done by
Hubble \& Humason 1931 \cite{hubble-h31}. They emphasized that
\emph{"The interpretation of red-shift as actual velocity, however,
does not command the same confidence, and the term "velocity" will be used
for the present in the sense of "apparent" velocity, without prejudice
as to its ultimate significance."}

Many years of detail studies of the linearity of the
redshift-distance law
was performed by Sandage at the Palomar 5m Hale telescope. Sandage developed a special program for 5m telescope
to discriminate between selected world models \cite{sandage61}.
One of the last paper of Sandage's team, devoted to analysis of the observed
redshift-distance relation, demonstrated linearity of $z(R)$ law
 in the interval of redshifts 
$ 0.001 \div 0.1$  \cite{sandage10}. 

Hence for the Local Universe we have observationally established the linear
redshift-distance Hubble-Humason-Sandage (HHS) law in the form:
\begin{equation}
\label{hhs-law}
z = \frac{H_{loc} R}{c} = \frac{V_{app}}{c}
\end{equation}
where $c$ is the velocity of light, $H_{loc}$ is the value of the Hubble
constant measured in the Local Universe, $R$ is the measured distance to
a galaxy, $V_{app} = H_{loc}\times R$ is the apparent radial velocity which
corresponds measured shift of spectral lines $z$: 
\begin{equation}
\label{z-law}
z = \frac{\lambda_{obs} - \lambda_{emit}}{\lambda_{emit}}
\end{equation}
where $\lambda_{obs}$ is the observed photon wavelength  at the telescope 
and $\lambda_{emit}$ is the wavelength of photon emitted at distance $R$.
The HHS law (\ref{hhs-law}) is also frequently called the Hubble law of redshifts.
Note that here $z$ is the cosmological part of the observed shift of spectral lines
after corrections for the solar system motions and averaging over peculiar velocities of galaxies.

The cosmological redshift is a universal physical phenomenon which does not depend
on the wavelength of a photon.
Very important cosmological question is about the minimal scale where the HHS law 
is true. Resent studies by Ekholm et al. 2001 \cite{ekholm01},  
Karachentsev et al. 2003 \cite{kara03} and Karachentsev et al. 2013 \cite{kara13}
demonstrated that according to modern data on 869 galaxy distances in the Local
Volume the linear Hubble law well established at small scales $1 \div 10$ Mpc. 
Remarkably, this is exactly the same
interval of scales where Hubble 1929 \cite{hubble29}
discovered the redshift-distance law with only 30 galaxies.

\begin{figure}[h]
\begin{center}
\includegraphics[width=9cm]{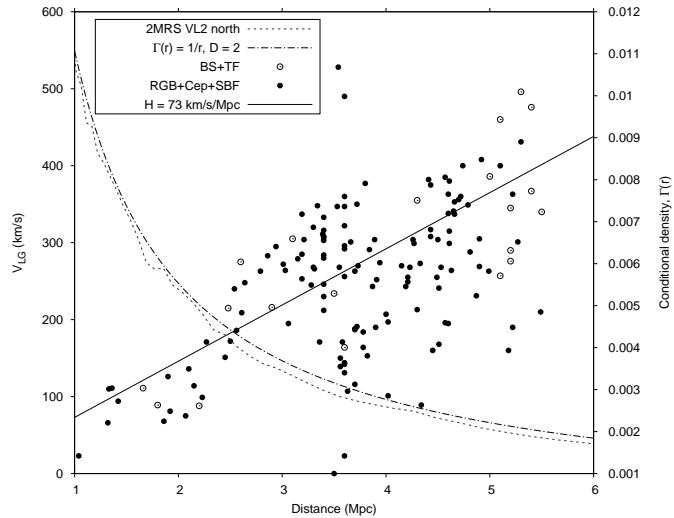}
\caption{\label{fig:1} 
Apparent radial velocity-distance relation $V_{app} =cz = H_{loc}R$
for 156 Local Volume galaxies is shown from \cite{kara03}. Also the density-radius relation $ \Gamma (r) \propto r^{-\gamma}$  from \cite{tekh-b16}
is shown dy thin lines for VL2N sample from 2MRS survey
\cite{huchra12} and for power-law density-radius relation for exponent
$\gamma = 1$.
}
\end{center}
\end{figure}

In Fig.1 apparent radial velocity-distance relation $V_{app} =cz = H_{loc}R$
for 156 Local Volume galaxies is shown from \cite{kara03}.
The value of the local Hubble constant is $H_{loc} = 72 \pm 3$ km/sec/Mpc,
which is consistent with recent estimations from different Local Universe surveys.

\section{Carpenter-Karachentsev-deVaucouleurs density-radius power-law}

The rich history of discovery and acute discussions around the density-radius relation
for the spatial galaxy distribution in the Local Universe  is presented in
\cite{baryshev-t02}, \cite{baryshev-t06}, \cite{baryshev-t12}, \cite{sylos11}, 
\cite{sylos14}.

Carpenter 1938 \cite{carpenter38} was the first who obtained from observations
of galaxy systems of different sizes the approximate power-law relation between
the number of galaxies $N$ in a cluster and the size $r$ of the clusters
in the form $N(r) \propto r^{1.5}$. 

Karachentsev 1966, 1968
\cite{kara66}, \cite{kara68} 
added an important aspect to Carpenter's result. 
He estimated average properties of 143 systems from binary galaxies to superclusters
and found evidence that both luminous and total (virial) mass densities are decreasing
with increasing size of a system. This showed for the first time that the mass–
radius behavior of the dark mass is also a power law, but the exponent can be different than for the luminous matter.

de Vaucouleurs 1970, 1971 \cite{devaucou70}, \cite{devaucou71} 
summarized his own and many others works in studies of galaxy systems from
pairs to superclusters, including clustering of Abel's rich galaxy clusters
\cite{abell58}, \cite{abell61}. Based on all available data
de Vaucouleurs made the decisive step in recognizing the cosmological significance
of the clustering of galaxies as the universal observational
power-law  density-radius relation \cite{devaucou70}. He considered this
fundamental cosmological law as the case for a hierarchical
cosmology.

Since that time the Carpenter-Karachentsev-deVaucouleurs (CKdeV) density-radius
empirical cosmological law was discovered and presented in the form
\begin{equation}
\label{rho-law}
\rho (r) = \rho_0 \,(r / r_0)^{-\gamma}
\end{equation}
where $\rho(r)$ is the mass density within a spherical volume of radius $r$
and $\rho_0$ and $r_0$ are the
density and radius at the lower cutoff of the structure. The available 
at that time galaxy data led to the power-law exponent $\gamma = 1.7$ .

Intriguingly, at international astronomical conferences, the Great Debate on the existence of very large scale structures in the observed
galaxy universe was originated. An acute discussion between homogeneity defenders
and inhomogeneity observers (see reviews \cite{baryshev-t06}, \cite{baryshev-t12})
is actually ongoing nowadays, though modern data demonstrate the existence
of galaxy structures with sizes up to 400 - 1000 Mpc
(e.g. \cite{shirokov16}).
The reason of the hot debates is that in the frame of the standard cosmological model
the homogeneous matter distribution is the basic mathematical assumption 
for the derivation of linear Hubble law of redshifts 
\cite{peebles93}, \cite{baryshev-t12}.

In fact, to understand the observed CKdeV density-radius relation
one needs to develop a new mathematical and physical concepts which
include discrete fractal stochastic structures. This was done by
Mandelbrot 1977 \cite{mandel77} in his theory of fractals, which opens
new perspective for description of complex discrete physical systems 
with properties very different from continues fluid flows.
Fractal approach to the analysis of the distribution of
galaxies was first used in \cite{mandel77}, \cite{pietro87}
and developed in \cite{sylos98}, \cite{gabr05}, \cite{sylos14}.  
For a detailed review of the history and prospects of
the fractal approach to the study of the large-scale distribution of galaxies
see ~\cite{baryshev-t06}, \cite{baryshev-t12}.

One of the most fundamental statistical properties of the general space
distribution of galaxies, which includes complex observed structures
(filaments, voids, shells, and walls), is the fractal dimension of the global
structure as a whole. 
According to \cite{gabr05} the fractal dimension $D$ of a stochastic fractal 
point process in 3-dimensional space can be inferred from the complete correlation function (conditional density) $\Gamma(r)$, which has the power-law: 
\begin{equation}\label{Gamma-def}
	\Gamma (r) =\frac{\langle n(\vec{r}_1) n(\vec{r}_2) \rangle}
                            {\langle n(r) \rangle}
                            = k \, r^{-\gamma} = k \, r^{-(3-D)}
\end{equation}
where $n(\vec{r}_i)$, is the particle number density inside volume $dV_i$
around point $i$ with the coordinates $\vec{r}_i$,  
$r = \mid\vec{r}_{12}\mid = \mid\vec{r}_{1} -\vec{r}_{2}\mid $, the vector of the distance between points 1 and 2, and $\langle
x \rangle$, the ensemble average of $x$. The second and third equalities are
written for isotropic stationary processes, where $D$ is the fractal dimension
and $\gamma = 3-D$ is called the co-dimension of the fractal.
The physical dimension of the $\Gamma(r)$ is 1/cm$^{3}$ and it is calculated under
the condition of all occupied points, this is why it is called the
conditional density. 

The power-law character of the conditional density (eq.\ref{Gamma-def}) is the
principal explanation of the CKdeV density-radius law (eq.\ref{rho-law}).
A more detailed analysis will include transition from number density $n(r)$
to mass density $\rho(r)$, which should also take into account the luminous
and dark matter. Fortunately, conditional density analysis of the real galaxy
catalogues shows that it is sufficient for  
describing the spatial galaxy distribution as a good first approximation. 

The statistical estimate of the complete correlation
function $\Gamma(r)$ (conditional density)
for the galaxy sample considered is defined as \cite{gabr05}:
\begin{equation}
	\Gamma(r) = \frac 1 {N_c(r)} \sum\limits_{i=1}^{N_c(r)} \frac {N_i(r)}{V(r)}
    \label{eq:gamma}
\end{equation}
where $N_i(r)$ is the number of points inside spherical volume $V(r)$ 
around {\it i}-th point and $N_c(r)$ is the
number of centers of test spheres, i.e., the number of points about which this
volume is circumscribed. It is important to bear in mind that averaging has to
be performed without going beyond the considered sample volume, and this
restriction has important effect on the value of the greatest available scale lengths. This condition strongly
restricts the scale-lengths
accessible for the analysis of galaxy correlations, because, strictly
speaking, in order to reliably compute the conditional density on some selected
scale, we must analyze much greater spherical region where all
test spheres are completely embedded.

For large galaxy redshift surveys the conditional density $\Gamma(r)$
is a directly determined quantity, which characterizes the
spatial, kinematical, and dynamical state of the Local Universe.
It can be
estimated from the power-law slope $\gamma$ (co-dimension of the fractal)
of the complete correlation function $\Gamma(r)$
without invoking any a priori assumptions about the evolution 
of non-baryonic dark
matter and its association with baryonic matter (galaxies) or the form of the
distribution of peculiar velocities of galaxies.

Note that the complete correlation function $\Gamma(r)$
has an important advantage
over reduced correlation function $\xi(r)$ (Peebles's two-point
correlation function) in that the computation
of conditional density requires no assumption about
the homogeneity of spatial galaxy distribution within
analyzed galaxy sample.

\begin{figure}[h]
\begin{center}
\includegraphics[width=9cm]{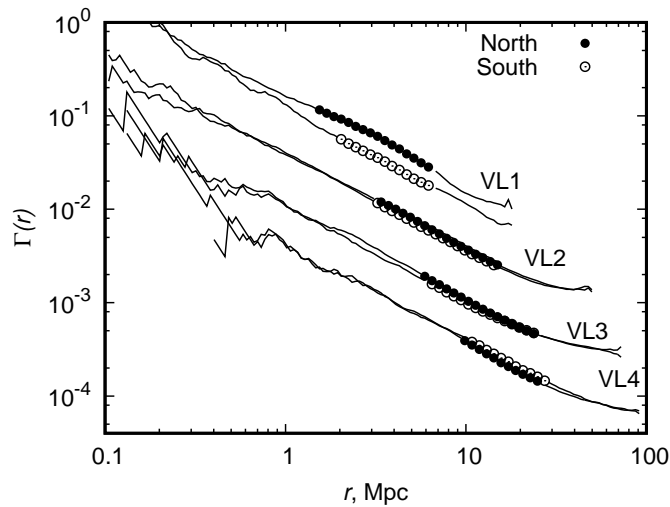}
\caption{\label{fig:1}
Conditional density for Volume Limited samples of 2MRS
galaxies in the Local Universe \cite{tekh-b16}. The large dots mark the conditional density values where the most reliable slope estimation is possible. The slope $\gamma = 1.0 \pm 0.1$  for all VL samples.
}
\end{center}
\end{figure}

Fig.2 shows the conditional density calculations
\cite{tekh-b16} for the largest complete all-sky galaxy redshift survey 2MRS
of the Local Universe \cite{huchra12}. 
The observed global space distribution of 2MRS galaxies 
can be described by the power-law complete
correlation function of the form $\Gamma(r) = k r^{-\gamma}$ with a
slope of $\gamma \approx 1$ over a wide interval of scale-lengths
spanning from 0.1 to 100 Mpc. 
The deeper all-sky volume limited sample is used (from VL1 to VL4)
then the larger is the maximum scale-length where the density power-law  
can be reliably estimated. The shift of the power-law maximum scale-length
is consistent with the stochastic
fractal model having the fractal dimension $D = 3-\gamma \approx  2$ 
in the whole interval of analyzed scales from 0.1 Mpc up to 100 Mpc.
\vspace{0.5cm}

In the frame of the LCDM theory of large scale structure formation
there are two important predictions:
\begin{itemize}
  \item the galaxy Local Universe is homogeneous after scales about 30 Mpc; 
  \item due to galaxy  peculiar velocities there is very large difference between slopes of conditional density calculated for redshift-based distances and real distances independent on $z$.
\end{itemize}
According to \cite{tekh-b16}, the Fig.3 shows results of the conditional density calculations for the Millennium
galaxy catalog (in a sample similar to S1VL2 2MRS) as a function of scale length 
in real and z space. The predicted slopes are very different for z- and r-space.
Also after scales about 30 Mpc the galaxy distribution becomes homogeneous. 

\begin{figure}[h]
\begin{center}
\includegraphics[width=9cm]{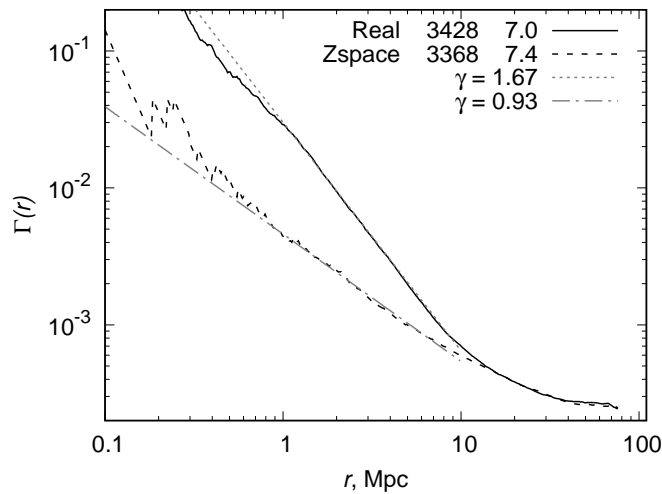}
\caption{\label{fig:1}
Conditional density of Millennium mock galaxy catalog in a sample
    similar to S1VL2 as a function of scale length in real and z space
    \cite{tekh-b16}. The slopes are estimated in the $1 < r < 10$~Mpc interval.
    After scales about 30 Mpc the mock galaxy distribution becomes homogeneous.}
\end{center}
\end{figure}

So for future testing of the nature of the Local Universe galaxy distribution
there are two possibilities - first, to get more deep all-sky galaxy redshift
surveys (at least up to 500 Mpc) and second, to compare conditional densities 
measured for redshift and real space: 
 $\Gamma(r_z) \Longleftrightarrow \Gamma(r_{real})$. 
Hence,
very important observational test of the large scale structure
origin in the Local Universe is
the direct measurements of the peculiar velocities of galaxies.
This will require further development of
redshift-independent methods for determining galaxy distances and performing
time consuming observational programs aimed to measurement of such
distances, like Cosmic Flows surveys ~\cite{tully13}.

Note that stochastic fractal structures naturally arise in physics as a result
of the dynamical evolution of complex systems. Physical fractals are discrete
stochastic systems characterized by power-law correlation functions. In
particular, fractal structures arise in turbulent flows and in
deterministic chaos of nonlinear dynamic systems.
Phase transitions and thermodynamics of self-gravitating
systems are also characterized by the formation of fractal
structures \cite{devega96}, \cite{devega98}, \cite{perdang90}. 
However, many important aspects of these
studies so far remain undiscovered.

\section{Physical interpretations of the relation between redshift and density laws}

Here I consider two possibilities for explanation of the surprising
coincidence of the observed spatial scales where two empirical
cosmological laws simultaneously  exit (see Fig.4)

\begin{figure}[h]
\begin{center}
\includegraphics[width=9cm]{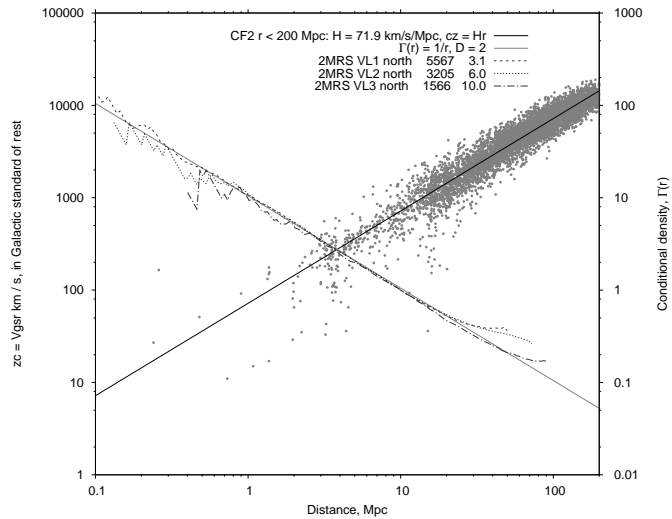}
\caption{\label{fig:1}
Demonstration of the Hubble-deVaucouleurs paradox in the Local Universe.
The Hubble-Humason-Sandage linear redshift law $cz=H_{loc}R$ and the fractal
Carpenter-Karachentsev-deVaucouleurs density law $\Gamma(r) = k r^{-\gamma}$
with $\gamma \approx 1$
coexist at the same length-scale interval $0.1 \div 100$ Mpc.
While in the frame of the SCM \textit{the linear redshift-distance relation is the strict consequence of homogeneity \cite{peebles91}}.}
\end{center}
\end{figure}

In the frame of the Friedmann model of the Standard Cosmological Model (SCM)
there is a deep paradox between  Hubble-Humason-Sandage 
linear redshift-distance law and 
Carpenter-Karachentsev-deVaucouleurs density-radius power-law.
This observational Hubble-deVaucouleurs (HdeV) paradox exists due to the very basis of SCM,
which explains the linear HHS law as a strict mathematical consequence of
the homogeneity of the matter distribution 
\cite{peebles91}, \cite{peebles93}, \cite{baryshev15}. 

For a solution of HdeV paradox within SCM one should assume a large amount
of homogeneously distributed non-baryonic dark matter and dark energy.
The dominance of homogeneous dark substance 
density over the usual baryonic matter (galaxies)
must start from scales where the linear HHS redshift-distance law
already exists. There are also several conceptual problems
with interpretation of space-expansion in SCM \cite{baryshev15}, 
\cite{francis07}, \cite{harrison95}, \cite{harrison93}.
\vspace{0.5cm}

Another solution of HdeV paradox can be obtained in the frame
of the Fractal Cosmological Model (FCM) \cite{baryshev08}, 
presented at the International conference
 \emph{Problems of Practical Cosmology 2008}.
In the frame of the FCM the space-geometry is static flat Minkowski space-time,
the gravitational interaction is described within Feynman's field gravity
approach \cite{feynman71}, \cite{feynman95}, \cite{baryshev-t12}, and 
the matter is dynamically evolving usual baryonic substance.

The spatial distribution of galaxies in the Local Universe is the stochastic
fractal structure with fractal dimension $D \approx 2$ and the cosmological
redshift is the new gravitational global effect due to the whole mass within
the sphere having radius equals to the distance between the source and
the observer. For fractal dimension $D = 2$ the mass of the sphere of radius
$r$ groughs as $M(r) \propto r^D \propto r^2$, hence the gravitational potential
is $\varphi \propto M/r \propto r^1$ and the cosmological global gravitational
redshift is the linear function of distance $z_{gl-gr} \propto r$.
It means that the surprising coincidence of length scales for both
HHS and CKdeV cosmological laws now is a natural prediction of the 
fractal cosmological model.

So an important task of Practical Cosmology is to observationally
distinct between expanding and static spaces, i.e. to establish
the nature of the observed cosmological redshift.
Note, that in the classical papers, Hubble 1929 \cite{hubble29} 
and Hubble \& Humason 1931 \cite{hubble-h31} 
emphasized that 
the cosmological part of the measured redshift 
should be called "\emph{apparent radial velocity}"
and actually can present the\emph{ de Sitter effect of
"slowing down of atomic vibrations"} - which is actually a kind of
the global gravitational effect. 
During all his life Hubble insisted on the 
necessity of the observational verification of the nature of the cosmological 
redshift and suggested several tests together with Tolman \cite{hubble35}.

Intriguingly, up to now, after 85 years of observational cosmology
there is no crucial experiment which
directly measure the real increasing distance with time.
In Sandage's List of 23 Astronomical Problems for the
1995 - 2025 years \cite{sandage95b} the first problem of the Practical
Cosmology is to test \emph{"Is the expansion real?"}.

The usually considered tests of space expansion
\begin{itemize}
  \item  Tolman's surface brightness $(1+z)^4$ test;
  \item  Time dilation with SN Ia $t(z) = t(0) (1+z)$;
  \item  CMBR temperature $T(z) = T(0)(1+z)$
\end{itemize}
can not distinct between space expansion redshift
and global gravitational redshift mechanisms.

The crucial  test of cosmological space expansion should measure
the real increasing distances with time.
Nowadays there are at least two proposals for such crucial tests of
the expansion of the Universe:
\begin{itemize}
  \item  Sandage's $z(t)$ test;
  \item  Kopeikin's  $\Delta \nu / \nu$ test in the solar system
\end{itemize}
It is important to note that on the verge of modern technology there are
possibilities for 
real direct observational tests of the physical nature of the cosmological redshift. First  crucial test of the reality of the space expansion was suggested by Sandage\cite{sandage62}, who noted that the observed redshift of a distant object (e.g. quasar) in expanding space must be changing with time according to relation
$dz/dt = (1+z)H_0 - H(z)$. In terms of radial velocity, the predicted change
$dv/dt \sim 1$ cm s$^{-1}$/yr. This may be within the reach of the future  ELT telescope \cite{pasquini05}, \cite{liske08}. In the case of the global gravitational
redshift the change of redshift equals zero.

Even within the Solar System there is a possibility to test the global expansion of the universe. According to recent papers by
Kopeikin\cite{kopeikin12,kopeikin15} the equations of light propagation used currently by Space Navigation Centers for fitting range and Doppler-tracking observations of celestial bodies contain some terms of the cosmological origin that are proportional to the Hubble constant $H_0$. Such  project as
PHARAO may be an excellent candidate for measuring the effect of the global cosmological expansion within Solar System, which
has a well-predicted blue-shift effect having magnitude
$\Delta\nu /\nu =2H_0 \Delta t \approx 4\times 10^{-15} (H_0/70kms^{-1}Mpc^{-1})
(\Delta t/10^3s)$, where $H_0$ is the Hubble constant, $\Delta t$ is the time of observations. In the case of the non-expanding Universe the frequency drift equals zero.

\section{Conclusion}

\emph{Cosmology at Small Scales} is very important part of astronomy. 
New mathematical and physical ideas in cosmology should be discussed
and tested by experiments and observations in the Local Universe
from the solar system scales up to the superclusters scales.

Surprises of recent observational cosmology of the Local Universe 
stimulate its further investigations.
A puzzling conclusion is that the Hubble's law, i.e. the strictly linear redshift-distance relation, is observed just inside strongly inhomogeneous galaxy distribution, i.e. deeply inside fractal structure at scales $1 \div 100$ Mpc.
This empirical fact presents a profound challenge to the standard cosmological model where the homogeneity is the basic explanation of the Hubble law, and \textit{"the connection between homogeneity and Hubble's law was the first success of the expanding world model"} (Peebles et al. 1991 \cite{peebles91}). However the spectacular 
observational fact (Fig.4) is that the Hubble's law is not a consequence of homogeneity
of the galaxy distribution, as it was assumed during almost the whole
history of cosmology. 

New type of global
physical laws can appear at cosmological scales which make cosmology especially
creative science. 
Intriguingly, up to now there is no crucial experiment which
directly measure the real increasing distance with time.
The global gravitational cosmological redshift can be such new
physical phenomenon which should be tested by observations and experiments.

New powerful mathematical methods of fractal structures analysis
should be developed for investigation of the large scale structure
of the Universe.
Even new approaches for description of gravitational interaction in the frame
of modern theoretical physics can be tested at all scales from solar
system up to the cosmological scales.

This is possible due to very fast development of observational technics and
theoretical models which is applied to astronomical objects.
Theoretical models utilize modern theoretical physics and even its
possible extensions, which can be tested by observations.
In conclusion we may say that now we are entering in the golden
age of cosmological physics of the Local Universe.
So the research in the field of \emph{Cosmology at Small Scales} is 
a perspective direction in modern physical science.

\section*{Acknowledgements}
I am grateful to  Pekka Teerikorpi, Georges Paturel,
Luciano Pietronero and Francesco Sylos Labini for many years of collaboration in study large scale structure of the Universe. Also I thank
Daniil Tekhanovich for joint work in analysis of  modern Local Universe galaxy catalogues.
This work has been supported by the Saint Petersburg State University (grant No.
6.38.18.2014).

\end{document}